\documentclass[12pt,modern]{aastex61}
\usepackage{graphics,graphicx}
\usepackage{hyperref}
\usepackage{amssymb}
\usepackage{amsmath}
\usepackage{comment}
\usepackage{grffile} 
\usepackage{empheq} 

\received{March 1, 2018}
\revised{April 18, 2018}
\accepted{\today}
\submitjournal{AAS journals.}

\shortauthors{Bouma, Masuda, Winn}
\shorttitle{Binarity and planet occurrence rates}

\renewcommand{\a}{_{\rm a}}
\newcommand{\s}{_{\rm s}}
\newcommand{\p}{_{\rm p}}

\renewcommand{\b}{_{\rm b}}
\begin{document}
    
\title{ Biases in Planet Occurrence Caused by Unresolved Binaries in
        Transit Surveys }
\correspondingauthor{L. Bouma}
\email{luke@astro.princeton.edu}
\author[0000-0002-0514-5538]{L. G. Bouma}
\affiliation{
    Department of Astrophysical Sciences,
    Princeton University,
    4 Ivy Lane, Princeton, NJ 08540, USA}
\author[0000-0003-1298-9699]{Kento Masuda}
\affiliation{
    Department of Astrophysical Sciences,
    Princeton University,
    4 Ivy Lane, Princeton, NJ 08540, USA}
\affiliation{
    NASA Sagan Fellow}
\author[0000-0002-4265-047X]{J. N. Winn}
\affiliation{
    Department of Astrophysical Sciences,
    Princeton University,
    4 Ivy Lane, Princeton, NJ 08540, USA}
\begin{abstract}
Wide-field surveys for transiting planets, such as the NASA {\it
Kepler} and {\it TESS} missions, are usually conducted without knowing
which stars have binary companions. Unresolved and unrecognized
binaries give rise to systematic errors in planet occurrence rates,
including misclassified planets and mistakes in completeness
corrections.  The individual errors can have different signs, making
it difficult to anticipate the net effect on inferred occurrence
rates. Here we use simplified models of signal-to-noise limited
transit surveys to try and clarify the situation. We derive a formula
for the apparent occurrence rate density measured by an observer who
falsely assumes all stars are single. The formula depends on the
binary fraction; the mass function of the secondary stars; and the
true occurrence of planets around primaries, secondaries, and single
stars. It also takes into account the Malmquist bias by which binaries
are over-represented in flux-limited samples. Application of the
formula to an idealized {\it Kepler}-like survey shows that for
planets larger than 2\,$R_\oplus$, the net systematic error is of
order 5\%.  In particular, unrecognized binaries are unlikely to be
the reason for the apparent discrepancies between hot Jupiter
occurrence rates measured in different surveys.  For smaller planets
the errors are potentially larger: the occurrence of Earth-sized
planets could be overestimated by as much as 50\%. We also show that
whenever high-resolution imaging reveals a transit host star to be a
binary, the planet is usually more likely to orbit the primary star
than the secondary star.
\end{abstract}
%
%


\section{Introduction}

One of the goals of exoplanetary science is to establish how common,
or rare, are planets of various types.  Knowledge of planet occurrence
rates is helpful for inspiring and testing theories of planet
formation, designing the next generation of planet-finding surveys,
and simply satisfying our curiosity.  One method for measuring
occurrence rates is to monitor the brightnesses of many stars over a
wide field, seeking evidence for planetary transits.  This was the
highest priority of the NASA {\it Kepler} mission.  Great strides have
been made in the analysis of {\it Kepler} data, including progress
towards measuring the fraction of Sun-like stars that harbor
Earth-like planets~\citep{
  youdin_exoplanet_2011,petigura_prevalence_2013,dong_fast_2013,
  foreman-mackey_exoplanet_2014,burke_terrestrial_2015}.

A lingering concern about these studies is that in most cases,
investigators have assumed that all of the targets in the survey are
single stars~\citep[\textit{e.g.},][]{
howard_planet_2012,fressin_false_2013,
dressing_occurrence_2015,burke_terrestrial_2015}.  In reality, many of
the sources that are monitored in a transit survey are unresolved
multiple-star systems, mainly binaries.  Unrecognized binaries cause
numerous systematic errors in the planetary occurrence rates.  For
example, when there is a transiting planet around a star in a binary,
the additive constant light from the second star reduces the
fractional loss of light due to the planet.  This makes transit
signals harder to detect and lowers the number of detections.  On the
other hand, a binary system presents two opportunities to detect
transiting planets, which could increase the overall number of
detections.

At the outset of this study it was not clear to us whether the neglect
of binaries is a serious problem, or even whether the net effect of
the errors is positive or negative.  The goals of this study were to
provide a framework for dealing with these issues, and to gauge at
least the order of magnitude of the systematic effects.  In this
spirit, our models are idealized.  We do not attempt a detailed
correction of the results from {\it Kepler} or any other real transit
survey.  We took inspiration from the insightful analytic model for
transit surveys by~\citet{pepper_using_2003}.

This paper is organized as follows.  The next section enumerates the
various errors that arise from unrecognized binaries.  Then in
Section~\ref{sec:simplest}, we develop an idealized model of a transit
survey in which all planets have identical properties, and all stars
are identical except that some fraction are in binary systems.  This
simple model motivates the derivation of a general formula, given in
Section~\ref{sec:general_formula}, that allows for more realistic
stellar and planetary populations.  We use this formula in
Section~\ref{sec:more_complicated} to explore more realistic models.
We discuss the errors due to unrecognized binaries for specific cases
of current interest: the occurrence of Earth-like planets; the
apparent discrepancy between hot Jupiter occurrence rates measured in
different surveys; and the shape of the ``evaporation valley'' in the
planet radius distribution that was brought to light by
\citet{fulton_california-_2017}.  We summarize and discuss all the
results in Section~\ref{sec:discussion}.

\section{Understanding the errors}
\label{sec:concept}

Imagine that a group of astronomers wants to measure the mean number
of planets per star.  They are particularly interested in planets of
radius $r$ and stars of mass $M$ and radius $R$.  They obtain a time
series of images of some region of the sky, and prepare light curves
for a large number of unresolved sources.  Then they search these
light curves for transit signals and detect all the signals for which
\begin{equation}
  \frac{\delta}{\sigma}
  >
  \left(\frac{{\rm S}}{{\rm N}}\right)_{\rm min}.
\label{eq:S_N_thresh}
\end{equation}
Here the signal, $\delta$, is the observed transit depth, the
dimensionless fraction by which the total light fades during transits.
Note that although $\delta$ is often equated with $(r/R)^2$, this is
not true when the host star is a member of an unresolved binary.  In
those cases, $\delta$ is smaller than $(r/R)^2$ because of the
constant light from the binary companion, an effect often called the
``dilution'' of the transit signal.  The noise, $\sigma$, is the
fractional uncertainty in the determination of the flux of the source,
which may include multiple stars that are blended together.  The
threshold signal-to-noise ratio depends on the desired level of
confidence that the signal is real.

The astronomers analyze their data assuming that all the sources are
single stars.  In particular they do not have accurate enough
parallaxes to tell that some of the stars appear to be overluminous.
They count the number $N_{\rm det}$ of transit signals that appear to
be produced by the desired type of planet around the desired type of
star.  They also count the number $N_\star$ of ``searchable stars'' in
their survey, i.e., the number of stars of the desired type that are
bright enough to have allowed for the detection of a transit signal
with amplitude $(r/R)^2$.  They estimate the occurrence rate to be
\begin{equation}
  \Lambda = \frac{N_{\rm det}}{N_\star}
                    \, \frac{1}{p_{\rm tra}},
  \label{eq:occ_rate_simple}
\end{equation}
where the geometric transit probability, $p_{\rm tra}$, accounts for
the fact that most planetary orbits are not aligned close enough with
our line of sight to produce transits.

There are many potential pitfalls in this calculation.  Some genuine
transit signals are missed even if they formally exceed the
signal-to-noise threshold, because of the probabilistic nature of
transit detection.  Planets can be misclassified due to statistical
and systematic errors in the catalogued properties of the stars.  Some
transit-like signals are spurious, arising from noise fluctuations or
failures of ``detrending'' the astrophysical or instrumental
variations in the photometric signal.  Poor angular resolution leads
to blends between eclipsing binary stars and other stars along nearly
the same line of sight, producing signals that mimic those of
transiting planets.

Here, though, we will focus exclusively on problems that arise from
the fact that many stars exist in unresolved binary systems.  We will
also focus on the errors in planet occurrence as a function of radius,
rather than orbital period. This is because when more than one transit
is detected (as is usually required by the surveyors), the orbital
periods can be measured without ambiguity regardless of whether the
host star is single or one member of a binary.  Even with this narrow
focus, there are numerous sources of error.  All three of the
quantities in Equation~\ref{eq:occ_rate_simple} are biased:
\begin{enumerate}
    \item The number of detected planets, $N_{\det}$, is actually the
      number of detected planets that {\it appear} to have size $r$,
      orbiting stars that {\it appear} to have mass $M$ \added{and
      radius $R$}.  Whenever the planet-hosting star is part of a
      binary,
    \begin{itemize}
        \item the planet's size could be misclassified because of the
          reduction in the amplitude of the photometric signal;
        \item the host star's properties could be misclassified
          because its light is combined with a second star of a
          different spectral type.
    \end{itemize}
    \item The number of searchable stars, $N_\star$, is biased
    \begin{itemize}
        \item toward lower values, because it does not include all of
          the secondary stars that were inadvertently searched for
          transiting planets;
        \item toward higher values, because some of the stars that
          appeared to be searchable are in fact binaries for which the
          amplitude of the photometric signal would have been reduced
          to an undetectable level.
    \end{itemize}
    \item The transit probability $p_{\rm tra}$ is biased because the
      planet-hosting star could be misclassified.  At fixed orbital
      period, the transit probability is proportional to
      $\rho^{-1/3}$, where $\rho$ is the stellar mean
      density~\citep{Winn2010}.  Therefore, any errors in determining
      the host star's mean density lead to errors in the correction
      for the transit probability.
\end{enumerate}

There are at least two other complications that may arise, which are
not represented in Equation~\ref{eq:occ_rate_simple}.  The first one
is an observational effect. Within the sample of apparently searchable
stars, the ratio between the number of binary and single stars will
differ from the ratio that would be found in a volume-limited sample.
This is due to a type of Malmquist bias.  The total luminosity of a
binary is larger than the luminosity of either the primary or
secondary star.  This means that for transit signals of a given
amplitude, sources that are binaries appear to be searchable at
greater distances from the Earth.  Binaries are therefore
over-represented in the collection of apparently searchable stars.

The other complication is astrophysical: the true occurrence rate of a
certain type of planet may depend on whether the host is a single
star, the primary star of a binary, or the secondary star of a binary.
The rate might also depend on the characteristics of the binary, such
as the mass ratio and orbital period.  Such differences could be
caused by the requirement for long-term dynamical stability, or
differences in the planet formation process.  When the search sample
includes both singles and binaries, the detected planets are thereby
drawn from different occurrence distributions~\citep[see][]{
  wang_occurrence_2015,kraus_impact_2016}.

Given all of the confusing and opposing sources of error, we will
proceed in stages.  We start with a model so simple that everything
can be written down on the back of a napkin, and build up to an
analytic model allowing for generality in the distribution of the
binaries and the planets they host.


\section{Simple models}
\label{sec:simplest}

\subsection{One type of star, one type of planet}
\label{sec:model_1}

Since binarity produces \deleted{the largest}\added{large} errors in inferred
planetary radii when the two stellar components are similar, we
begin by considering a universe in which
all stars are identical, with mass $M$, radius $R$, and luminosity
$L$.  Single stars are uniformly distributed in space with a number
density of $n\s$ stars per cubic parsec, and binary systems are
uniformly distributed with number density $n\b$.  In this scenario,
stars are never misclassified because the combined light of a binary
has the same color and spectrum as a single star.  We further assume
that all planets have the same radius, $r$, and occur around single
stars and members of binaries at the same rate, $\Lambda_r$.

Our naive observers conduct a transit survey. To calculate the
occurrence rate of planets with radius $r$, they count the number of
detections of signals with amplitude $(r/R)^2$.  Then they identify
all the sources that appear to have been searchable for a signal of
amplitude $(r/R)^2$.  Here and throughout the rest of this paper, we
assume that the limiting source of noise is the photon-counting noise
from the source, i.e., $\sigma \propto 1/\sqrt{F}$, where $F$ is the
total flux of the source.  Thus the observers determine the minimum
$F_0$ for which detection would have been possible, and count the
number of sources with $F>F_0$. This will include all the single stars
out to a maximum distance
\begin{equation}
  d_0 = \sqrt{\frac{L}{4\pi F_0}}.
  \label{eq:dmax}
\end{equation}
Since binaries are twice as luminous, the condition $F>F_0$ will
include binaries out to the larger distance of $d_0\sqrt{2}$.  None of
the stars in binaries will appear to have a planet of radius $r$,
because of the dilution of the transit signal.  Thus, the {\it
apparent} occurrence rate $\Lambda\a$ of planets of radius $r$ is
\begin{equation}
    \Lambda_{{\rm{a}},r} = 
        \frac{\Lambda_r \, n\s d_0^3}
        {n\s d_0^3 + n\b (d_0\sqrt{2})^3}.
    \label{eq:rate_at_rp}
\end{equation}
This apparent rate is smaller than the true rate by a factor
\begin{equation}
    \frac{\Lambda_{{\rm{a}},r}}{\Lambda_r} = 
        \frac{1}{1 + 2^{3/2}(n\b/n\s)}.
    \label{eq:correction_rp}
\end{equation}

\begin{figure}[!tb]
    \begin{center}
        \includegraphics[width=0.6\textwidth]{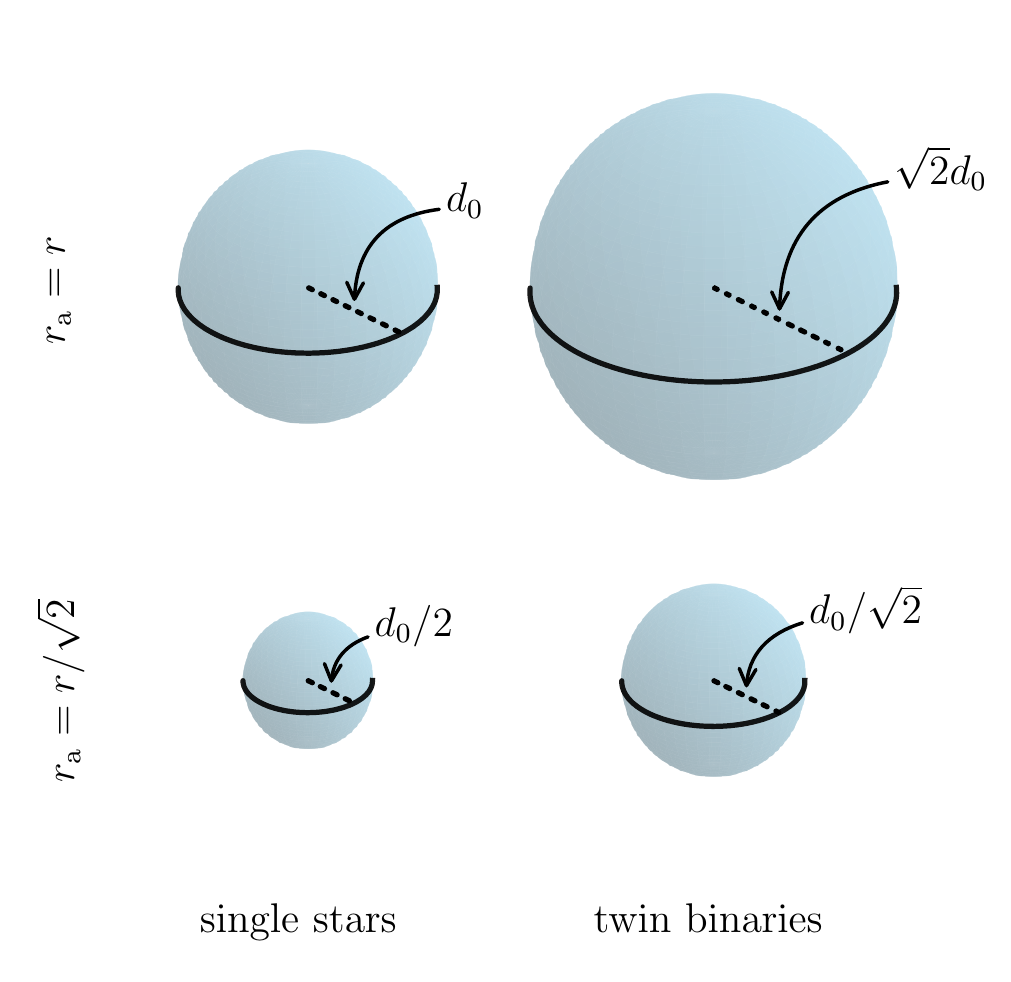}
    \end{center}
    \caption{ {\it Top.}---Volumes within which sources appear to be
    searchable for planets of radius $r$.  Single stars are searchable
    out to a distance $d_0$, at which point they become too faint to
    allow the detection of transits.  Binaries are brighter and
    therefore appear to be searchable out to a larger distance.  {\it
    Bottom.}---Volumes within which stars appear to be searchable for
    planets of radius $r/\sqrt{2}$.  }
    \label{fig:model_1_volumes}
\end{figure}

The observers will also detect some transiting planets around stars
with binary companions.  The amplitude of these signals is
\begin{equation}
  \frac{L(r/R)^2}{L + L} = \frac{1}{2}\left(\frac{r}{R}\right)^2 =
  \left(\frac{r/\sqrt{2}}{R}\right)^2,
    \label{eq:delta_obs_general} 
\end{equation}
leading the astronomers to believe they have discovered a population
of planets with radius $r/\sqrt{2}$.  To calculate the corresponding
occurrence rate, they count the stars for which this type of signal
would have been detectable.  The limiting flux for detection in this
case is $4F_0$, because the signal amplitude has been reduced by a
factor of two and the noise level must also be reduced by a factor of
two.  The condition $F>4F_0$ is met for single stars within a distance
$d_0/2$, and binaries within a distance $d_0\sqrt{2}/2$.  Therefore,
the observers will calculate the occurrence rate of this new type of
planet to be
\begin{equation}
    \Lambda_{{\rm{a}},r/\sqrt{2}} = 
        \frac{2\, \Lambda_r \, n\b (d_0\sqrt{2}/2)^3}
        {n\s (d_0/2)^3 + n\b (d_0\sqrt{2}/2)^3}
    =
    \frac{2\, \Lambda_r \cdot2^{3/2}(n\b/n\s)}{1 + 2^{3/2} (n\b/n\s)}.
    \label{eq:correction_diluted_rp}
\end{equation}
Figure~\ref{fig:model_1_volumes} illustrates the
volumes enclosing the apparently searchable single and binary stars.

\begin{figure}[!tb]
    \begin{center}
        \includegraphics[width=.6\textwidth]{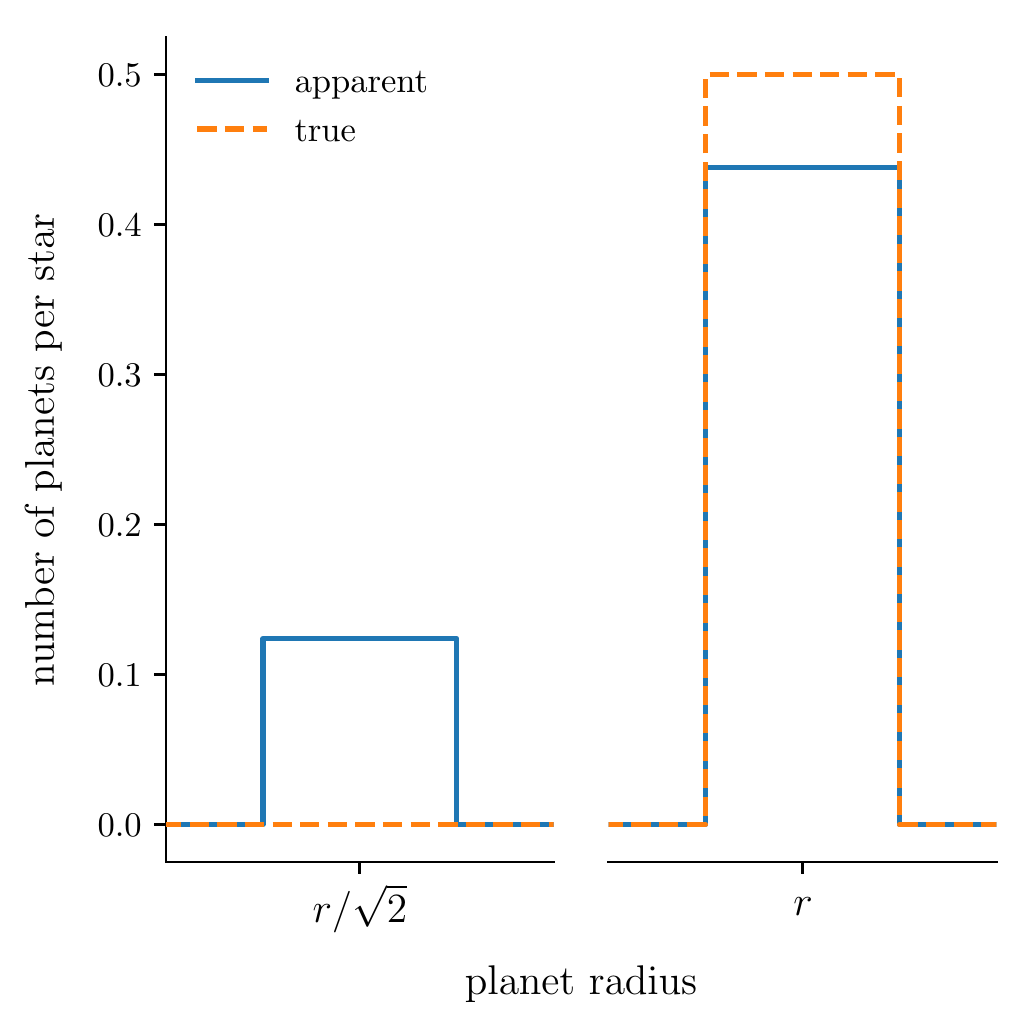}
    \end{center}
    \vspace{-0.5cm}
    \caption{ Apparent occurrence rate $\Lambda\a$, and true
    occurrence rate $\Lambda$ in a universe with only one type of
    star, one type of planet (with radius $r$), and $n\b/n\s=0.05$,
    for $n\b$ the number density of binary systems, and $n\s$ the
    number density of single stars.  The occurrence rate of planets
    with radius $r$ is underestimated, and the occurrence rate of
    planets with radius $r/\sqrt{2}$ is overestimated.}
    \label{fig:occ_rate_model_1}
\end{figure}

We can now assess the severity of the errors, for a given value of the
binary-to-single ratio $n\b/n\s$. For stars with masses from 0.7 to
1.3~$M_\odot$, \citet{raghavan_survey_2010} found the multiplicity
fraction~--~the fraction of systems in a volume-limited sample that
are multiple~--~to be 0.44.  Assuming all multiple systems are
binaries, this gives a binary fraction
\begin{equation}
  \frac{n\b}{n\s + n\b} \approx 0.44,
\end{equation}
which implies $n\b/n\s \approx 0.79$.  Of course not all of these
binaries are ``twin'' binaries as we have assumed in our simple
calculation.  Later, in Section~\ref{sec:general_formula}, we will
allow for a continuum of properties for the secondary stars. For now,
we might guess that only a tenth of the binaries have pairs of stars
close enough in brightness to produce errors as significant as those
we have been considering.  Adopting the value $n\b/n\s\approx0.05$, we
find
\begin{equation}
    \frac{\Lambda_{{\rm{a}},r}}{\Lambda_r} = 0.88,~~
    \frac{\Lambda_{{\rm{a}},r/\sqrt{2}}}{\Lambda_r} = 0.25.
    \label{eq:ratios}
\end{equation}
\added{The apparent and true rates are illustrated in 
Figure~\ref{fig:occ_rate_model_1}.}
All together, the various effects produce biases of order 10\% in the
occurrence rates.  We will see that this level of error is
characteristic of many of our more complicated models as well.

\subsection{Planets of different sizes}
\label{sec:model_1_density}

We will now generalize to allow for a continuum of planet sizes.   We
introduce the occurrence rate density $\Gamma$, the number of planets
per star per unit planet radius, so that
\begin{equation}
    \Gamma(r) {\rm d}r \equiv \frac{ {\rm d}N_{\rm det} }{N_\star}
            \, \frac{1}{p_{\rm tra}}.
    \label{eq:rate_density_defn}
\end{equation}
The naive observers are measuring $\Gamma\a(r\a)$, the apparent
occurrence rate density of planets with an ``apparent radius''
\begin{equation}
  r\a = R \sqrt{\delta}.
\end{equation}

For a given value of $r\a$, the observers assemble the sample of
$N_\star(r\a)$ sources for which a signal of amplitude $\delta =
(r\a/R)^2$ could have been detected.  We showed in the previous
section that when all stars are identical, this sample will contain
binaries and single stars in the ratio $2^{3/2}(n\b/n\s)$.  Here we
will denote this ratio by $\mu$, which will be calculated later in a
more general context.  With this definition, the number of single
stars is proportional to $1/(1+\mu)$ and the number of binaries is
proportional to $\mu/(1+\mu)$.

From within this sample of apparently searchable sources, the
observers sum the number of detections ${\rm d}N_{\rm det}$ of planets
with an apparent radius between $r\a$ and $r\a + dr\a$.  There will be
three contributions to this sum,
\begin{align}
  {\rm d}N_{{\rm det},0} &= 
      \frac{N_\star(r\a) p_{\rm tra}}{1+\mu}
      \Gamma_0(r\a)\mathrm{d}r\a,
  \label{eq:n0} \\
  {\rm d}N_{{\rm det},1} &= 
      \frac{\mu N_\star(r\a) p_{\rm tra}}{1+\mu} 
      \Gamma_1(\sqrt{2}r\a)\mathrm{d}(\sqrt{2}r\a),
	\label{eq:n1} \\
  {\rm d}N_{{\rm det},2} &=
      \frac{\mu N_\star(r\a) p_{\rm tra}}{1+\mu} 
      \Gamma_2(\sqrt{2}r\a)\mathrm{d}(\sqrt{2}r\a).
	\label{eq:n2}
\end{align}
where here and elsewhere, the subscript 0 refers to single stars, 1
refers to primary stars of binaries, and 2 refers to secondary stars
of binaries. The apparent occurrence rate density is therefore
\begin{equation}
  \Gamma\a(r\a) = \frac{{\rm d}N_{\rm det}}{{\rm d}r\a} \frac{1}{N_\star(r\a) p_{\rm tra}} =
  \frac{\Gamma_0(r\a)}{1+\mu} +
  \frac{\mu\sqrt{2}\,\Gamma_1(\sqrt{2}r\a)}{1+\mu} +
  \frac{\mu\sqrt{2}\,\Gamma_2(\sqrt{2}r\a)}{1+\mu}.
	\label{eq:formula_twin_binary}
\end{equation}
Our earlier results, Equations~\ref{eq:rate_at_rp}
and~\ref{eq:correction_diluted_rp}, are recovered by integrating this
formula over $r$ after letting the respective rate densities be
\begin{equation}
  \Gamma_i(r) = \Lambda_{r\p}\, \hat{\delta}(r-r\p),~~{\rm for}\ i\in\{0,1,2\}
\end{equation}
where $\hat{\delta}$ is the Dirac delta function, and $\Lambda_{r\p}$
is the number of planets per star with size $r\p$ considered in
Section~\ref{sec:model_1}.

\subsection{Binaries with different mass ratios}
\label{sec:general_formula}

Next we generalize to allow for a spectrum of different properties for
the secondary stars.  For simplicity we assume that the stars form a
one-parameter family specified by the stellar mass $M$.  This is
approximately the case for main-sequence stars.  The functions $L(M)$
and $R(M)$ give the luminosity and radius as a function of mass, and
$f(q)$ is the distribution of binary mass ratios in a volume-limited
sample.  We also assume that observers perceive all the binaries to be
isolated stars with the same mass as the primary star, i.e., the light
from the secondary star is either too faint or too similar to the
primary star to make a difference in the stellar classification.
Finally, for clarity of presentation we make the simplifying
assumption that $\Gamma_0$, $\Gamma_1$ and $\Gamma_2$ do not depend on
stellar mass, although our formalism can easily accommodate such a
dependence.

To compute the apparent occurrence rate density, we need to make
the following modifications to Equation~\ref{eq:formula_twin_binary}:
\begin{enumerate}
\item The Malmquist bias is different.  For a given value of $r\a$, the
  number of binary systems in the searchable sample with mass ratio
  $(q, q+\mathrm{d}q)$ is
\begin{equation}
    {N_\star(r\a)\over{1+\mu}}\, {n_{\rm b}\over n_{\rm s}}
    \left[{L(M) + L(qM) \over L(M)}\right]^{3/2}
    f(q)\,\mathrm{d}q,
    \label{eq:N_binaries}
\end{equation}
where $\mu$ is given by
\begin{equation}
    \mu= 
    \int_0^1 {n_{\rm b}\over n_{\rm s}}\left[{L(M) + L(qM)
    \over L(M)}\right]^{3/2} f(q)\,\mathrm{d}q.
    \label{eq:mu_general}
\end{equation}
Equation~\ref{eq:N_binaries} replaces ${\mu N_\star(r\a)/(1+\mu)}$
in Equations~\ref{eq:n1} and~\ref{eq:n2}.
The whole equation is then integrated \added{over $q$} to give the
number of detected planets.

\item The apparent radius of a planet in a binary system now depends
on whether the host star is the primary or the secondary star. When the host
is the primary star, we write $r=\mathcal{D}_1 r\a$, where
\begin{equation}
    \mathcal{D}_1
    = \left[ \frac{ L(M) + L(qM) }{ L(M) } \right]^{1/2}
    \label{eq:dil_1}
\end{equation}
is the appropriate dilution factor.
When the host is the secondary star, a correction must also be made
to account for the different radius of the secondary star.
In that case $r=\mathcal{D}_2 r\a$, where
\begin{equation}
  \mathcal{D}_2
  = \frac{R(qM)}{R(M)}
    \left[ \frac{L(M) + L(qM)}{L(qM)} \right]^{1/2}.
  \label{eq:dil_2}
\end{equation}

\item When a transiting planet is detected around a secondary star,
  the naive observers make the wrong correction for the transit
  probability.  At a fixed orbital period, $p_{\rm tra}$
  in Equation~\ref{eq:n2} must be multiplied by a factor of
\begin{equation}
  {R(qM) \over R(M)}q^{-1/3}.
\end{equation}

\end{enumerate}

Taking these modifications into account, a general formula for the
apparent rate density is
\begin{align}
    \notag
    \Gamma\a(r\a) &= \frac{1}{1+\mu} \,
    \left\{ \Gamma_0(r\a)+ 
    \frac{n\b}{n\s}
    \left[ \int_0^1 \mathrm{d}q \,
           \mathcal{D}_1^3 f(q)\cdot
           \mathcal{D}_1 \Gamma_1(\mathcal{D}_1r\a)\,
    \right.   
    \right. \\
    & \quad\quad\quad\quad\quad \left.\left.
    +\int_0^1 {\rm d}q\, 
         \mathcal{D}_1^3 f(q)\cdot \mathcal{D}_2
         \Gamma_2(\mathcal{D}_2r\a)\cdot
         \frac{R(qM)}{R(M)} q^{-1/3}
    \right] \right\}.
    \label{eq:general_Gamma_a}
\end{align}
We note that the combination $\mathcal{D}_1^3f(q)$ is the mass-ratio
distribution for binaries contained within the searchable sample of
sources.  With this in mind we may rewrite
Equation~\ref{eq:general_Gamma_a} as
\begin{align}
    \Gamma\a(r\a)
    =
    {1\over 1+\mu} \,
    \left[
       \Gamma_0(r\a)
       +\mu\,
       \left\langle
       \mathcal{D}_1 \, \Gamma_1\left(\mathcal{D}_1r\a\right)
       +
       {\mathcal{D}_2} \, \Gamma_2\left(\mathcal{D}_2r\a\right)
       \,
       {R(qM) \over R(M)}q^{-1/3}
       \right\rangle
    \right],
\end{align}
where the angle brackets denote averaging over all the binaries in the
searchable sample.


\section{Case Studies}
\label{sec:more_complicated}

\begin{deluxetable}{cccc}
    



\caption{Table of contents to case studies in 
Section~\ref{sec:more_complicated}.}
\label{tab:case_study_toc}

\tablenum{1}

\tablehead{
  \colhead{Section} &
  \colhead{True rate density} &
  \colhead{Apparent rate density} & 
  \colhead{Applicable where?}
}

\startdata
  \ref{sec:model_2} & power-law & Eqs.~\ref{eq:power_law_correction}
  \&~\ref{eq:powerlaw_vary_binary} & 2-17\,$R_\oplus$; no gap; no HJs
  \\
  \ref{sec:model_3} \& \ref{subsec:whichstar} & broken power-law &
  Fig.~\ref{fig:occ_rate_model_3_log} & 0-17\,$R_\oplus$; 
  no gap; no HJs
  \\
  \ref{sec:further_models} & broken power-law + gap & Fig.~\ref{fig:model_4} & 
  0-17\,$R_\oplus$; has gap; no HJs
  \\
  \ref{sec:hjs} & gaussian & Fig.~\ref{fig:gaussian_HJ} & only hot Jupiters
  \\
\enddata


\tablecomments{
    Aside from varying the true rate density, within each case study we also 
    vary the number of planets orbiting secondary stars, relative to the 
    number orbiting single stars. We use the following notation throughout.
    $\alpha$---power-law exponent in stellar mass-luminosity relation, 
    $L\propto M^\alpha$;
    $\beta$---power-law exponent in volume-limited binary mass-ratio 
    distribution, $f(q)\propto q^\beta$;
    $\gamma$---power-law exponent in true rate density above 2\,$R_\oplus$, 
    $\Gamma(r)\propto r^\gamma$.
}

\end{deluxetable}

We can now gauge the size of the systematic errors associated with
unresolved binaries, for \deleted{cases of interest.}\added{the cases
of interest listed in Table~\ref{tab:case_study_toc}.} Throughout this
section we assume $R\propto M$, and $L \propto M^{\alpha}$ where
$\alpha = 3.5$, as is roughly the case for main-sequence stars.  With
these assumptions, Equations~\ref{eq:dil_1} and~\ref{eq:dil_2} become
\begin{equation}
  \mathcal{D}_1
  =  (1+q^\alpha)^{1/2}~~{\rm and}~~\mathcal{D}_2
  = q(1+q^{-\alpha})^{1/2}.
\end{equation}

\subsection{Power-law planet radius distribution}
\label{sec:model_2}

Based on {\it Kepler} data, \citet{howard_planet_2012} found the
radius distribution of planets between 2 and 17~$R_\oplus$ orbiting
Sun-like stars within 0.25\, AU to be consistent with a power law,
\begin{equation}
  \Gamma_0(r) \propto r^\gamma,
\label{eqn:power_law}
\end{equation}
with $\gamma=-2.92\pm 0.11$.  Their analysis ignored the effects of
binarity.  We can use our formalism to estimate the resulting level of
systematic error.

To warm up we will again consider the case in which all stars are
identical.  We assume further that the distributions of planets around
primaries and secondaries differ only by multiplicative factors,
\begin{equation}
    \Gamma_1 = Z_1 \Gamma_0,~~\Gamma_2 = Z_2 \Gamma_0.
\end{equation}
Application of Equation~\ref{eq:general_Gamma_a} gives
\begin{equation}
    \frac{\Gamma\a(r\a)}{\Gamma_0(r\a)} = 
    \frac{1}{1+\mu}
    +
    2^{\frac{\gamma+1}{2} } \frac{\mu}{1+\mu} \left(Z_1 + Z_2\right),
    \label{eq:model5_apparent_rate_density}
\end{equation}
where $\mu = 2^{3/2} n\b/n\s$.  The ratio does not depend on $r\a$ and
thus, under our assumptions, the effect of twin binaries is simply to
change the normalization of the radius distribution.  If we further
assume that planet occurrence is independent of system multiplicity,
i.e., $Z_1=Z_2=1$, then
\begin{equation}
    \frac{\Gamma\a(r\a)}{\Gamma_0(r\a)} = 
    \frac{1 + 2^{\frac{\gamma+3}{2}}\mu}{1 + \mu}.
    \label{eq:power_law_correction}
\end{equation}
Adopting $\gamma=-2.92$, and a twin binary fraction $n\b/n\s=0.05$, as
before, we obtain $\mu = 0.14$ and $\Gamma\a/\Gamma_0 = 1.003$.  The
correction is tiny because, by coincidence, the reported value of
$\gamma$ is very close to $-3$, the value for which
Equation~\ref{eq:power_law_correction} reduces to unity and the
effects of binarity cancel out completely.  This suggests that
although \citet{howard_planet_2012} derived the occurrence rate
density under the false assumption that all stars are single, the
resulting systematic error is negligible.

\begin{figure}[!tb]
    \centering
    \includegraphics[width=0.6\textwidth]{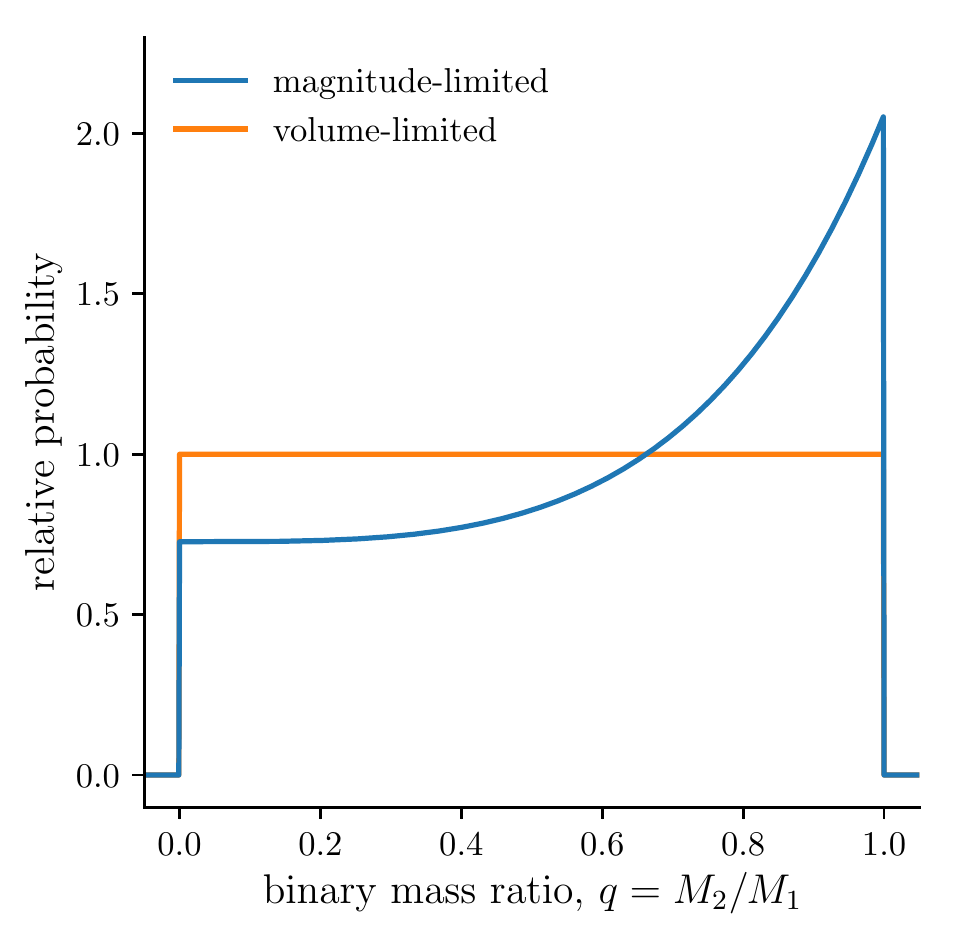}
    \caption{ The mass ratio distribution of binaries in a
    magnitude-limited sample, assuming the underlying volume-limited
    distribution is uniform and the luminosity-mass relation is
    $L\propto M^{3.5}$. For transit signals of a given amplitude, the
    sample of searchable sources is magnitude-limited, causing high
    mass ratios to be over-represented. }
    \label{fig:q_distribn_mag_limited}
\end{figure}

For a more accurate analysis we now consider a distribution of binary
mass ratios.  Studies of binaries in the local neighborhood suggest
that the distribution of mass ratios is nearly uniform between zero
and unity~\citep{raghavan_survey_2010}.  We consider the more general
possibility of a power-law dependence, $f(q) = \mathcal{N}_q q^\beta$,
where $\mathcal{N}_q$ is a normalization constant.  In this case
Equation~\ref{eq:mu_general} gives
\begin{equation}
    \mu = \frac{1}{\mathcal{N}_q} \frac{n\b}{n\s}
    \int_0^1 (1+q^\alpha)^{3/2}q^\beta\,{\rm d}q.
    \label{eq:mu_power_law_q}
\end{equation}
Figure~\ref{fig:q_distribn_mag_limited} illustrates the Malmquist
bias for the particular case of a flat distribution.

Again assuming $\Gamma_0=\Gamma_1=\Gamma_2$, the apparent occurrence
rate density is 
\begin{align}
    \notag
    \frac{\Gamma\a(r\a)}{\Gamma_0(r\a)} 
    &=
    \frac{1}{1+\mu}
    \left[
      1 + \frac{1}{\mathcal{N}_q} \frac{n\b}{n\s}
    \left(
      \int_0^1 {\rm d}q\,q^\beta (1+q^\alpha)^{\frac{\gamma+4}{2}} +
      \right.
      \right. \\
      &\quad\quad\quad\quad\quad\quad\quad\quad\quad\quad\quad
      \left.\left.
      \int_0^1 {\rm d}q\,q^{\beta+\gamma+\frac{5}{3}} 
      (1+q^\alpha)^{\frac{3}{2}}
      (1+q^{-\alpha})^{\frac{\gamma+1}{2}}
    \right)
    \right],
    \label{eq:powerlaw_vary_binary}
\end{align}
Adopting the realistic numerical values $n\b/n\s=0.79$, $\alpha =
3.5$, $\beta=0$, and $\gamma=-2.92$, the summed integrals in
Equation~\ref{eq:powerlaw_vary_binary} give $(\ldots)\approx 1.503$,
and $\Gamma\a/\Gamma_0 = 1.048$.  The apparent occurrence rate density
is 4.8\% higher than the true value.
\added{Exploring different values for the power-law index,
we find that $\Gamma\a/\Gamma_0$ lies between $0.99$ and 
$1.10$, for $\gamma=-3.5$ and $\gamma=-2.5$, respectively.}
We conclude that for planets around Sun-like stars with orbits within
0.25\,AU and radii in the range 2--17\,$R_\oplus$, the systematic
errors associated with binarity are on the order of a few percent.

\subsection{Broken power-law planet radius distribution}
\label{sec:model_3}

\begin{figure}[!t]
    \centering
    \epsscale{1.15}
    \plottwo{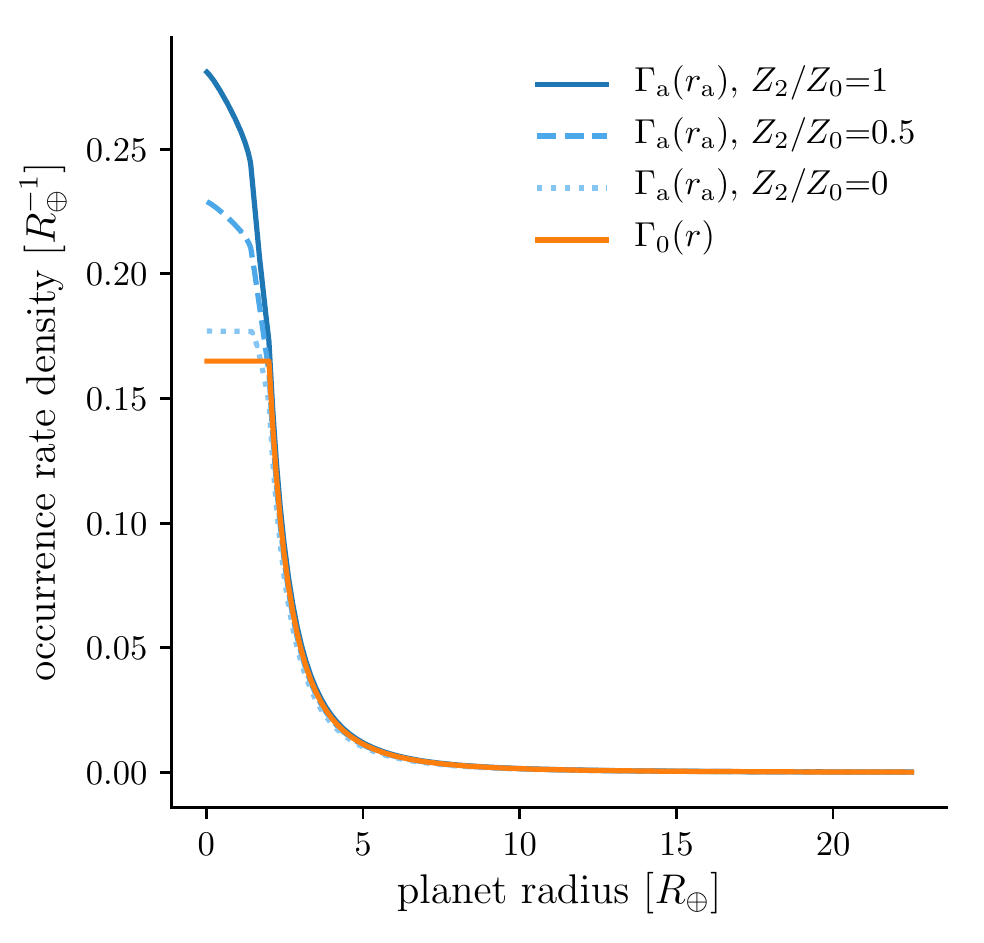}{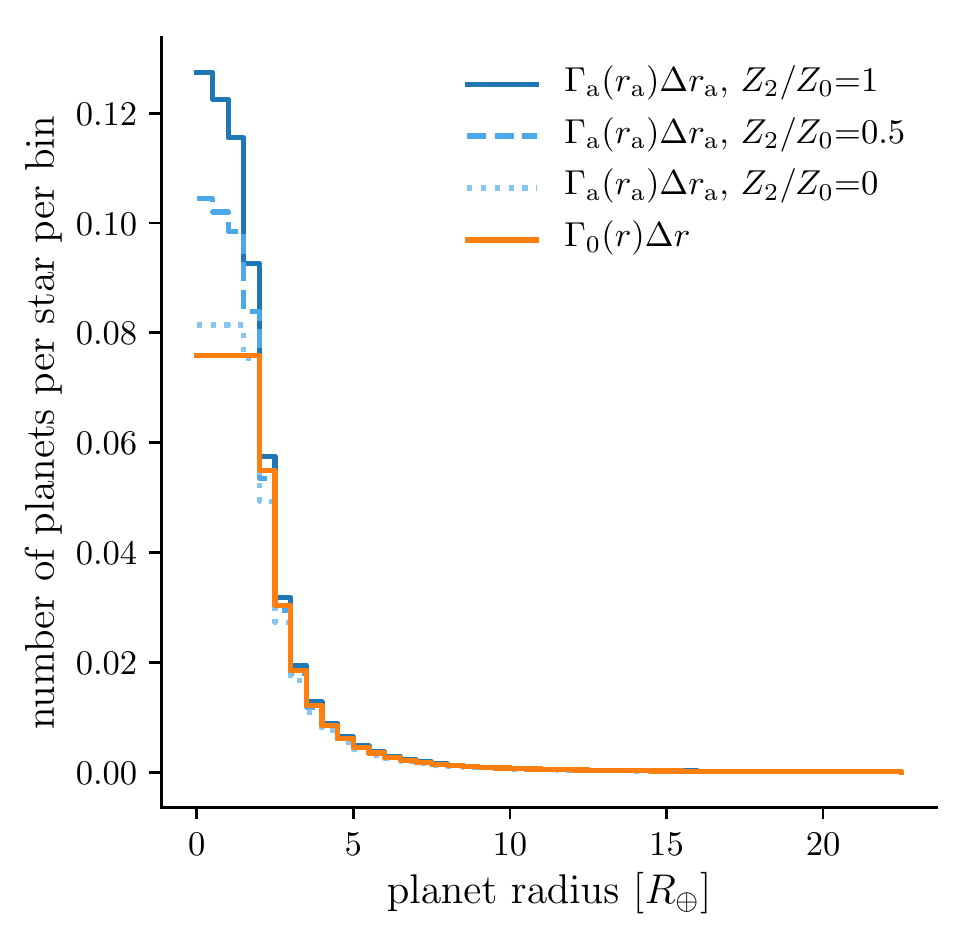}
    \caption{ {\it Left.}---Apparent occurrence rate density
      ($\Gamma\a$), compared to the true occurrence rate density
      ($\Gamma_0$) for single stars and primary stars in binaries.
      Three different cases are plotted, with different choices for
      the number of planets per secondary star relative to
      that for single stars ($Z_2/Z_0$).
      {\it Right.}---Same, integrated over $0.5\,R_\oplus$
      bins to give apparent and true occurrence rates.
      In both cases the true planet
      radius distribution is specified by
      Equation~\ref{eq:model3_radius_distribution}.
    }
    \label{fig:occ_rate_model_3_log}
\end{figure}

For planets smaller than 2\,$R_\oplus$, the true occurrence rate
density is more uncertain because such planets are more difficult to
detect.  To investigate the systematic errors associated with binarity
on the inferred occurrence rate of Earth-sized planets, we need to
make plausible assumptions about the true occurrence rate density.
Obviously if the rate density continues to vary as $r^{-3}$ down to
much smaller planet sizes, the results of the preceding section will
hold.  But there is no particular reason to think this will be the
case and indeed, some investigators have concluded that the rate
density begins to level off to a constant value as $r$ decreases below
$2R_\oplus$~\citep{petigura_plateau_2013}.  To investigate the
implications we consider a broken power-law:
\begin{align}
    f(r)
    &\propto
    \left.
    \begin{cases}
        r^\gamma & \text{for } r\geq 2R_\oplus \\
        {\rm constant} & \text{for } r\leq2R_\oplus.
    \end{cases}
    \right.
    \label{eq:model3_radius_distribution}
\end{align}
In this case the integrals that appear in
Equation~\ref{eq:general_Gamma_a} are tedious to work out
analytically, leading us to evaluate them numerically.\footnote{We
refer the interested reader to our online code for performing this
computation: \url{github.com/lgbouma/binary_biases}, commit
\texttt{fd41b4b}.} Figure~\ref{fig:occ_rate_model_3_log} shows the
results, again for the case $n\b/n\s=0.79$, $\alpha = 3.5$, $\beta=0$,
and $\gamma=-2.92$.

The most obvious aspect of the results is that the occurrence rate of
planets smaller than the breakpoint radius of 2\,$R_\oplus$ is now
substantially overestimated.  If secondary stars have the same planet
population as the single stars, then the occurrence rate density of
small planets is overestimated by $50\%$.  The magnitude of the
systematic error decreases if there are fewer planets around
secondaries; if secondaries host no planets at all, the occurrence
rate is overestimated by only $10\%$.

The effects on larger planets are not as substantial.  To be
quantitative we consider the occurrence rate of giant planets with
$r>8\,R_\oplus$, by integrating the occurrence rate density displayed
in Figure~\ref{fig:occ_rate_model_3_log}:
\begin{equation}
    \Lambda_{\rm giant,a} =
      \int_{8\,R_\oplus}^{\infty} \Gamma\a(r\a)\,{\rm d}r\a,~~{\rm
      and}~~
    \Lambda_{\rm giant,0} =
      \int_{8\,R_\oplus}^{\infty} \Gamma_0(r)\,{\rm d}r.
\end{equation}
In this case the occurrence rate density is underestimated.  The
difference between the apparent and true rates is largest when the
secondary stars do not host any planets, and has a magnitude
$\Lambda_{{\rm giant},0}/\Lambda_{\rm giant,a} = 1.13$.
If instead the secondary stars have half as many giant planets as single
stars, then the factor is reduced from 1.13 to 1.06.

\subsection{Which star has the planet?}
\label{subsec:whichstar}
\begin{figure}[!t]
    \centering
    \includegraphics[width=\textwidth]{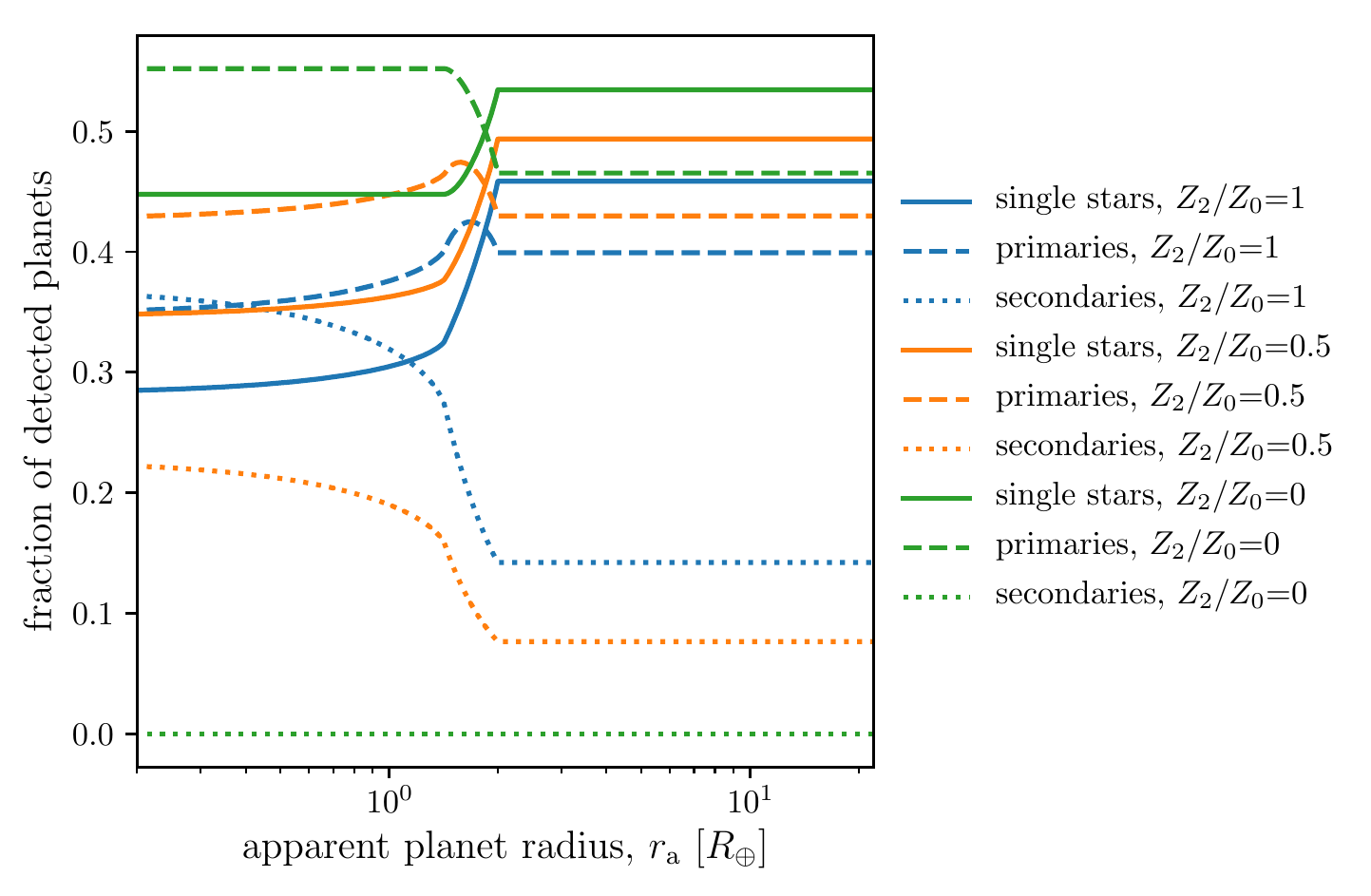}
    \caption{ Fraction of detected planets that orbit single stars
      (solid lines), primary stars of binaries (dashed lines), and
      secondary stars of binaries (dotted lines), as a function of the
      apparent radius of the planet.  We assume that single stars and
      primaries have the same number of planets per star.  Three
      different cases are plotted, differing in the relative
      occurrence of planets around secondary stars ($Z_2/Z_0$).  The
      true radius distribution is given by
      Equation~\ref{eq:model3_radius_distribution}.}
    \label{fig:frac_model_3}
\end{figure}

Our formalism also provides a way to calculate the relative
probability that a detected planet orbits a single star, the primary
star, or a secondary star.  This information can be used to help
interpret the results of a transit survey.  It is also relevant to the
fairly common situation in which transits are detected from a source,
and subsequent high-resolution imaging reveals the source to be a
multiple-star system.  The question then arises: which star is the
host of the detected planet?  Without further observations the answer
is often unclear.  To illustrate we return to the scenario described
in the previous section: a broken power-law radius distribution with
$\gamma=-2.92$, along with the choices $n\b/n\s=0.79$, $\alpha = 3.5$,
and $\beta=0$.  We also try different assumptions for the rate of
planets around secondaries, i.e., different choices for $Z_2/Z_0$.
Dividing each component of Equation~\ref{eq:general_Gamma_a} by the
total rate density, we get the fraction of detections from single
stars, primaries, and secondaries as a function of apparent radius.

Figure~\ref{fig:frac_model_3} shows the results.  When the apparent
radius exceeds $2\,R_\oplus$, and planets exist about secondaries at
the same rate as the primary star ($Z_2/Z_0=1$), then the planet is
$3$ times more likely to orbit the primary star than the secondary
star.  If secondaries host half as many planets as primaries
($Z_2/Z_0=0.5$), then a detected planet is $5$ times more likely to
orbit the primary star.

The situation for smaller apparent radii is more nuanced.  For
$Z_2/Z_0=0$ or 0.5, any transit signals from binaries are still always
more likely to arise from the primary star.  However, if planets exist
at the same rate in primaries and secondaries ($Z_2/Z_0=1$), then
below apparent radii of $0.4\,R_\oplus$, more of the detected planets
in binaries come from secondaries.  They are actually much larger
planets for which the transit signal amplitude has been substantially
reduced by the light from the primary star.  As we consider apparent
radii ranging from 2 to 1.4\,$R_\oplus$, the fraction of detected
planets with binary companions increases by anywhere from
$6\%$~to~$12\%$, depending on the relative occurrence of planets about
primaries and secondaries.

\subsection{A gap in the radius distribution}
\label{sec:further_models}

\begin{figure}[!t]
    \centering
    \epsscale{1.15}
    \plottwo{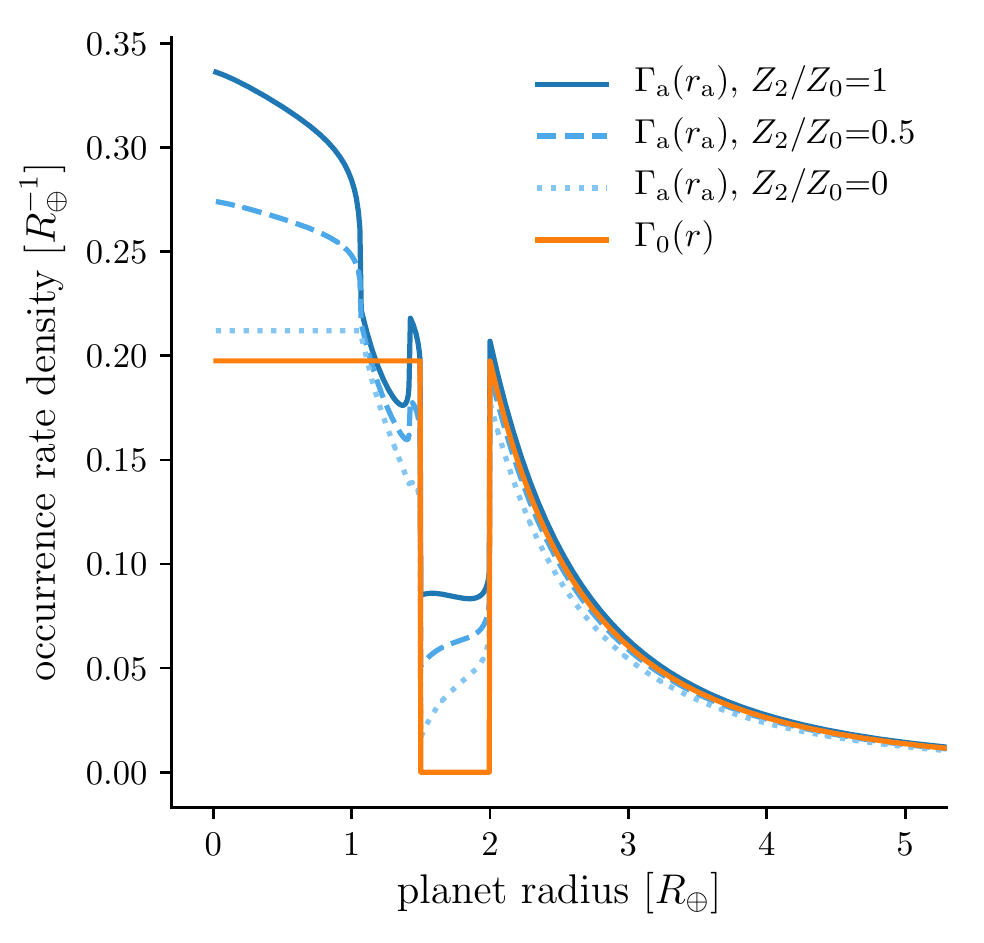}{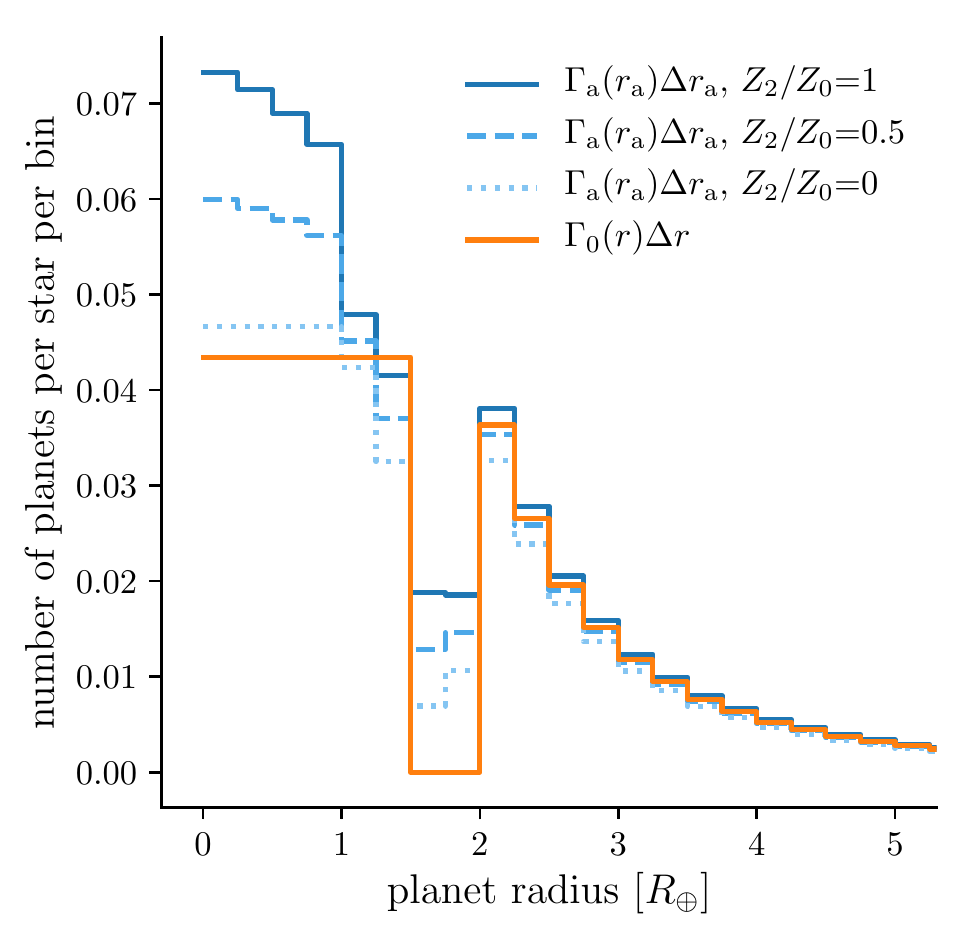}
    \caption{ {\it Left.}---True and apparent occurrence rate densities,
      and {\it right.}---occurrence rates, as a function of planet
      radius, using a model for the radius distribution that exhibits
      a gap between 1.5 and 2\,$R_\oplus$
      (Equation~\ref{eq:model4_radius_distribution}). 
      \added{The peak in apparent occurrence rate density between 1.3 and 
      1.5$\,R_\oplus$ (solid dark blue line, relative to dotted light blue 
      line) is the contribution from secondaries with true radii greater
      than 2\,$R_\oplus$.}
    }
    \label{fig:model_4}
\end{figure}

\citet{fulton_california-_2017} recently reported a ``gap'' in the
radius distribution of close-in planets around Sun-like stars, between
planet sizes of 1.5 and 2\,$R_\oplus$.  This can also be visualized as
a ``valley'' in the occurrence rate density as a function of radius
and period.  The existence of the gap has been independently
corroborated from a sample of KOIs with asteroseismically-determined
stellar parameters~\citep{van_eylen_asteroseismic_2017}.  Such a
feature had been predicted as a consequence of the gradual
photo-evaporation of the hydrogen-helium atmospheres of small rocky
planets, during the first 100~Myr of the age of the system when the
host star produces a higher flux of ultraviolet and X-ray
radiation\added{~\citep[\textit{e.g.},][]{lopez_role_2013,owen_kepler_2013,owen_evaporation_2017}}.

The occurrence rate calculations that led to this discovery \deleted{did not
carefully account for}\added{gave only a preliminary accounting for} 
unresolved binaries.
Intuitively we expect
unresolved binaries to have a blurring effect, filling in any gaps in
the \added{true} radius distribution and making them appear less empty than in
reality.  To investigate the quantitative effects we make identical
assumptions as in Section~\ref{sec:model_3}, except that \deleted{the radius
distribution is assumed to have}\added{in analogy with
\citet{fulton_california-_2017}'s findings we assume the true radius 
distribution has} a complete absence of planets with
sizes between 1.5 and 2\,$R_\oplus$:
\begin{align}
    f(r)
    &\propto
    \left.
    \begin{cases}
        r^\gamma & \text{for } r\geq 2\,R_\oplus, \\
        0 & \text{for } 1.5\,R_\oplus < r < 2\,R_\oplus, \\
        {\rm constant} & \text{for } r\leq 1.5\,R_\oplus.
    \end{cases}
    \right.
    \label{eq:model4_radius_distribution}
\end{align}
Figure~\ref{fig:model_4} shows the resulting true and apparent
occurrence rate densities.  For the case $Z_2/Z_0=1$, unresolved
binaries reduce the contrast of the gap by nearly a factor of two,
while also producing spurious features for apparent sizes beneath
1.5\,$R_\oplus$.

\added{The quantitative results of this numerical experiment obviously
depend on the input assumptions.  The more general point is that
whenever there is a gap-like feature in the true rate density
distribution, the effects of binarity will tend to fill it in.  If a
gap is observed in an apparent distribution, the true distribution
must have a gap at least as deep, if not deeper.}  Thus, the gap
identified by \citet{fulton_california-_2017} may be even more devoid
of planets than it appears.


\subsection{Gaussian radius distribution}
\label{sec:hjs}

\begin{figure}[!tb]
    \centering
    \includegraphics[width=.6\textwidth]{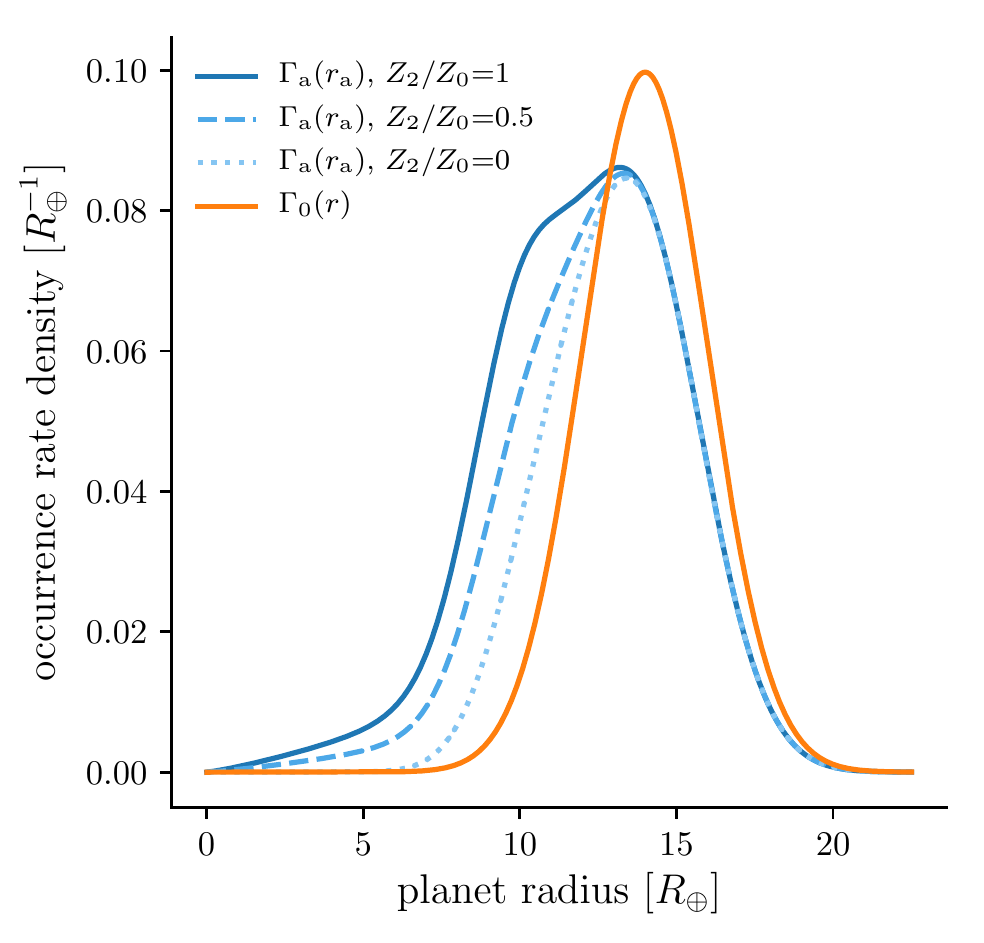}
    \caption{ True and apparent occurrence rate densities, for a
    population of planets with true radii $r$ drawn from a Gaussian
    distribution with mean $14\,R_\oplus$ and standard deviation
    $2\,R_\oplus$.  This is similar to the hot Jupiter distribution
    presented by~\citet{petigura_CKS_2017}.  }
    \label{fig:gaussian_HJ}
\end{figure}

The planet population may include some special members with a distinct
radius distribution.  For example in the recent study
by~\citet{petigura_CKS_2017}, hot Jupiters appear as an island in
period-radius space, rather than as a component of a power-law
distribution extending to smaller planets and longer periods.  For
this reason we test the effects of unresolved binaries on a Gaussian
radius distribution,
\begin{equation}
    f(r) = 
    \frac{1}{\sqrt{2 \pi \sigma_r^2}} \,
    \exp \left[ -\frac{(r-\bar{r})^2}{2\sigma_r^2} \right],
\end{equation}
with $\bar{r} = 14\,R_\oplus$ and $\sigma_r = 2\,R_\oplus$.  As before,
we allow for different normalizations of the planet populations around
single, primary, and secondary stars: $\Gamma_i(r) = Z_i f(r)$.
Figure~\ref{fig:gaussian_HJ} shows the results for the apparent radius
distribution, $\Gamma\a(r\a)$.  As one would expect, the effect is to
smear the radius distribution toward smaller values, because some of
the hot Jupiters now appear to be smaller planets.  Less obviously,
the integrated rate of hot Jupiters is also affected.  We calculate
the apparent hot Jupiter rate $\Lambda_{\rm HJ,a}$ by integrating the
apparent rate density for $r\a > 8\,R_\oplus$.  If secondaries host
any planets at all, the apparent rate is {\it greater} than the true
rate.  For instance, when hot Jupiters are just as common around
secondary stars as single or primary stars, then $\Lambda_{\rm
HJ,a}/\Lambda_{{\rm HJ},0} = 1.23$.
The apparent rate is higher because a larger number of stars were
searched than are accounted for in the rate calculation.

\section{Summary and Discussion}
\label{sec:discussion}

Ignoring binarity introduces systematic errors to star and planet
counts in transit surveys, which in turn lead to biases in the derived
planet occurrence rates.  Thus far, occurrence rate calculations using
transit survey data have mostly ignored stellar
multiplicity~\citep[\textit{e.g.},][]{howard_planet_2012,fressin_false_2013,foreman-mackey_exoplanet_2014,dressing_occurrence_2015,burke_terrestrial_2015}.
We do not claim to have solved this problem. Our aim was to clarify
the various effects and assess the size of the errors in various
situations. We hope that our formalism will also be useful to those
investigators who do attempt more realistic calculations of occurrence
rates.

The calculations presented in Section~\ref{sec:model_2} suggest that
for planets around Sun-like stars with periods shorter than a few
months and apparent radii exceeding $2\,R_\oplus$, binarity can
basically be ignored.  The errors are only of order a few percent.
For apparent radii below $2\,R_\oplus$, nature may not be as
forgiving.  The size of the effect depends on the true radius
distribution and in particular whether it continues to vary as
$\sim$$r^{-3}$ down to 1\,$R_\oplus$ and below.  The calculations
presented in Section~\ref{sec:model_3} suggest that the apparent rates
could be overestimated by as much as 50\%.

\paragraph{Earth-like planets} For the specific problem of calculating
the occurrence rate $\eta_\oplus$ of Earth-like planets with periods
of order one year, even a 50\% systematic error would not be the
dominant source of uncertainty at the moment.  Estimates by
\citet{youdin_exoplanet_2011},~\citet{petigura_prevalence_2013},~\citet{dong_fast_2013},~\citet{foreman-mackey_exoplanet_2014},
and~\citet{burke_terrestrial_2015} have found values of $\eta_\oplus$
ranging from $0.03$ to unity, with the wide variance arising from
small-number statistics as well as a host of other systematic
errors~\citep[see][Figure~17]{burke_terrestrial_2015}.  Rapid progress
is expected soon, though, and binarity will merit closer attention in
the future.  It may also be worth investigating the possible effects
of any dependence of occurrence rate densities on the period of the
binary.  Binaries with separations $\lesssim 10\,{\rm AU}$ could
provoke dynamical instabilities, leading to fewer Earth-like planets
per
star~\citep[\textit{e.g.},][]{holman_long-term_1999,wang_influence_2014,
kraus_impact_2016}.  This would affect transit survey measurements of
$\eta_\oplus$ beyond our rough estimate.

\paragraph{Hot Jupiters} Another frequently discussed problem
involving occurrence rates is the discrepancy between the occurrence
rates of hot Jupiters that have been measured in different surveys, as
summarized in Table~\ref{tab:hj_rates}.  In particular, the California
Planet Search (CPS), a Doppler survey, found a higher rate ($12\pm 2$
per thousand stars) than the {\it Kepler} transit survey
($5.7^{+1.4}_{-1.2}$).  It has been suggested that unresolved binaries
in the {\it Kepler} sample might be an important factor, but the
calculations presented in 
Section~\added{\ref{sec:hjs}}\deleted{\ref{sec:further_models}} suggest
that this is not the case.  We found that the effects of unresolved
binaries are likely to result in an overestimation of the hot-Jupiter
occurrence rate, which would worsen the discrepancy, rather than
relieve it. Some other possibilities are errors in stellar
classification of the {\it Kepler} stars, as well as differences in
the metallicity of the stars in each survey.  The occurrence rate of
hot Jupiters is known to vary strongly with stellar metallicity, with
$\Lambda \propto 10^{3.4{\rm [M/H]}}$ according to a recent
study~\citep{petigura_CKS_2017}.  \citet{guo_metallicity_2017} found
the mean metallicity of the {\it Kepler} stars to be [M/H]~$=-0.045\pm
0.009$, which is only slightly lower than that of the stars in the
Doppler survey, $-0.005\pm 0.006$.  Nevertheless, given the strong
dependence of occurrence on metallicity, even a difference of
$0.04$~dex could account for a factor of $10^{3.4(0.04)} = 1.37$
in the occurrence rate.  After correcting for metallicity, the
discrepancy between the {\it Kepler} and CPS occurrence rates is
reduced to the 1.7$\sigma$ level.

\paragraph{Radius corrections} Another application of our formalism is
to help in those cases in which it is not clear whether the primary or
secondary star is the source of a transit signal.  By this point
almost all of the {\it Kepler}\ Objects of Interest have been examined
with high resolution imaging~\citep{
  howell_speckle_2011,adams_adaptive_2012,adams_adaptive_2013,horch_observations_2012,
  horch_most_2014,lillo-box_multiplicity_2012,lillo-box_high-resolution_2014,dressing_adaptive_2014,
  law_robotic_2014,cartier_revision_2015,everett_high-resolution_2015,gilliland_hubble_2015,
  wang_influence_2015,wang_influence_2015-1,baranec_robo-ao_2016,ziegler_robo-ao_2017}.
The results of these programs have been summarized
by~\citet{furlan_kepler_2017}. Based on this information,
\citet{hirsch_assessing_2017} re-assessed the radius measurements of
the {\it Kepler} planets in binaries. They found that planet radii are
generally underestimated by a factor $r/r\a = 1.17$ if all planets
orbit primaries, and $r/r\a = 1.65$ if detected planets are equally
likely to orbit primaries and \deleted{singles}\added{secondaries}.
Our formalism provides a way to assess the relative probabilities that
a detected planet orbits the primary or secondary star.  The
calculations presented in \S\,\ref{subsec:whichstar} suggest that in
almost all cases the planet most likely orbits the primary star.

\added{\paragraph{Caveats \& outlook} In this paper we have focused
  exclusively on systematic errors in the planet radius distribution.
  Unresolved binaries will also have other effects on the inferred
  properties of a distribution of transiting planets.  For example,
  the combined light of a binary might lead to a mistaken estimate of
  the mass of the host star.  This would have implications for the
  occurrence rate as a function of orbital distance and insolation,
  since the orbital distance is usually calculated from the orbital
  period and the stellar mass.}

\added{Of course, many problems associated with binarity would
disappear if we had perfect knowledge about the companions to all
the stars in a transit survey. Progress toward this ideal is likely
to come soon, thanks to the combination of data from the ESA {\it
Gaia} astrometric mission \citep{Gaia2016} and the NASA {\it TESS}
mission~\citep{ricker_transiting_2014}.  The {\it TESS} transit
survey will concentrate on stars that are brighter than {\it Kepler}
stars by a factor of 30--100, and closer to the Earth by a factor of
5--10.  This will simplify all of the follow-up observations to
search for companions and measure their properties, with direct
imaging, Doppler monitoring, and astrometric monitoring.  Meanwhile
{\it Gaia} will deliver parallaxes, astrometic monitoring data, and
space-based apparent magnitudes for all of the stars relevant for {\it
TESS}, all of which will make it easier to identify binaries and other
multiple star systems.}

\acknowledgements{
It was a pleasure discussing this study with T.~Barclay, W.~Bhatti,
J.~Christiansen, F.~Dai, and T.~Morton.  This work made use of NASA's
Astrophysics Data System Bibliographic Services.
This work was performed in part under contract with the California Institute 
of Technology (Caltech)/Jet Propulsion Laboratory (JPL) funded by NASA through 
the Sagan Fellowship Program executed by the NASA Exoplanet Science Institute.
\newline
\software{\texttt{numpy}~\citep{walt_numpy_2011}, 
\texttt{scipy}~\citep{jones_scipy_2001}, 
\texttt{matplotlib}~\citep{hunter_matplotlib_2007}, 
\texttt{pandas}~\citep{mckinney-proc-scipy-2010},
\texttt{IPython}~\citep{perez_2007}
}
}


\newpage


\begin{deluxetable}{cccc}
    



\caption{Occurrence rates of hot Jupiters (HJs) about FGK dwarfs, as measured 
by radial velocity and transit surveys.}
\label{tab:hj_rates}

\tablenum{2}

\tablehead{\colhead{Reference} & \colhead{HJs per thousand stars} & 
\colhead{HJ Definition} 
} 

\startdata
\citet{marcy_observed_2005}
    & 12$\pm$2 & $a<0.1\,{\rm AU}; P\lesssim10\,{\rm days}$ \\
\citet{cumming_keck_2008} & 15$\pm$6 & -- \\
\citet{mayor_harps_2011} & 8.9$\pm$3.6 & -- \\
\citet{wright_frequency_2012} & 12.0$\pm$3.8 & -- \\
\citet{gould_frequency_2006} & $3.1^{+4.3}_{-1.8}$ & $P<5\,{\rm days}$ \\
\citet{bayliss_frequency_2011} & $10^{+27}_{-8}$ & $P<10\,{\rm days}$ \\
\citet{howard_planet_2012} 
    & 4$\pm$1 & $P<10\,{\rm days}; r_p=8-32R_\oplus$; solar 
    subset\tablenotemark{a} 
\\
-- & 5$\pm$1 & solar subset extended to $Kp<16$ \\
-- & 7.6$\pm$1.3 & solar subset extended to $r_p>5.6R_\oplus$. \\
\citet{moutou_corot_2013} 
    & 10$\pm$3 & {\it CoRoT} average; $P\lesssim 10\,{\rm days}$, 
    $r_p>4R_\oplus$  \\
\citet{petigura_CKS_2017} & $5.7^{+1.4}_{-1.2}$ &
    $r_p=8-24R_\oplus$; $P=1-10\,{\rm days}$; CKS stars\tablenotemark{b} \\
Santerne et al. (2018, in prep) & 9.5$\pm$2.6 & {\it CoRoT} galactic center \\
-- & 11.2$\pm$3.1 & {\it CoRoT} anti-center \\
\enddata


\tablecomments{
    The first four studies use data from radial velocity surveys; the rest
    are based on transit surveys. Many of these surveys selected different 
    stellar samples. ``--'' denotes ``same as row above''.
}
\tablenotetext{a}{
    \citet{howard_planet_2012}'s ``solar subset'' was defined as {\it 
    Kepler}-observed stars with $4100\,{\rm K}<T_{\rm eff}<6100\,{\rm K}$, $Kp 
    <15$, $4.0 < \log g < 4.9$. They required signal to noise $>10$ for planet 
    detection.
    }
\tablenotetext{b}{
    \citet{petigura_CKS_2017}'s planet sample includes all KOIs with 
    $Kp<14.2$, with a statistically insignificant number of fainter stars with 
    HZ planets and 
    multiple transiting planets.
    Their stellar sample begins with \citet{mathur_revised_2017}'s catalog of 
    199991 {\it Kepler}-observed stars.
    Successive cuts are: $Kp<14.2\,{\rm mag}$, $T_{\rm eff}=4700-6500\,{\rm 
    K}$, and $\log g = 3.9-5.0\,{\rm dex}$, leaving $33020$ stars.
}   
\end{deluxetable}


\newpage
\bibliographystyle{yahapj}                            
\bibliography{bibliography} 

\listofchanges
\end{document}